\documentclass[oldversion,a4paper]{aa}
\usepackage{txfonts}
\usepackage{natbib}
\usepackage{graphicx}

\bibstyle{aa}

\newcommand{\fifps}[2]{\centering\resizebox{#1}{!}{\includegraphics{#2}}}

\newcommand{\filps}[2]{\resizebox{#1}{!}{\rotatebox{-90}{\includegraphics{#2}}}}

\def\llm{{\sc LLModels}}
\def\synthm{{\sc SynthM}}
\def\atlas{{\sc ATLAS9}}
\def\tef{T_{\rm eff}}

\def\taucont{\tau_{5000}}

\def\paper1{{Paper~I}}

\begin{document}

\title{Stellar model atmospheres with magnetic line blanketing}
\subtitle{II. Introduction of polarized radiative transfer}
\titlerunning{Stellar model atmospheres with magnetic line blanketing. II.}

\author{S. A. Khan\inst{1,2} \and D. V. Shulyak\inst{2,3}}

\offprints{S. A. Khan, \email{skhan@astro.uwo.ca}}

\institute{%
Physics and Astronomy Department, University of Western Ontario, London, ON, N6A 3K7, Canada \and
Institut f\"ur Astronomie, Universit\"at Wien, T\"urkenschanzstra{\ss}e 17, 1180 Wien, Austria \and
Tavrian National University, Vernadskogo 4, 95007 Simferopol, Crimea, Ukraine }

\date{Received / Accepted}

\abstract{
The technique of model atmosphere calculation for magnetic Ap and Bp stars with polarized radiative transfer and magnetic line blanketing is presented. A grid of model atmospheres of A and B stars are computed. These calculations are based on direct treatment of the opacities due to the bound-bound transitions that ensures an accurate and detailed description of the line absorption and anomalous Zeeman splitting. The set of model atmospheres was calculated for the field strengths between 1 and 40\,kG. The high-resolution energy distribution, photometric colors and the hydrogen Balmer line profiles are computed for magnetic stars
with different metallicities and are compared to those of non-magnetic reference models and to the previous paper of this series.
The results of modelling confirmed the main outcomes of the previous study: energy redistribution from UV to the visual region and flux depression at 5200\,\AA. However, we found that effects of enhanced line blanketing when transfer for polarized radiation takes place are smaller in comparison to those obtained in our first paper where polarized radiative transfer was neglected. Also we found that the peculiar photometric parameter $\Delta a$ is not able to clearly distinguish stellar atmospheres with abundances other than solar, and is less sensitive than ${\Delta(V_1-G)}$ or $Z$ to a magnetic field for low effective temperature (${\tef=8000}$\,K). Moreover we found that the back determination of the fundamental stellar atmosphere parameters using synthetic Str\"omgren photometry does not result in significant errors.
}

\maketitle

\section{Introduction}

The magnetic line blanketing in stellar atmospheres due to the Zeeman effect is supposed to be responsible for observable characteristic features of magnetic chemically peculiar (CP) stars, for instance flux depressions \citep{kodaira} in the visual spectrum, and flux redistribution from UV region to the visual \citep{leckrone1}. A review of literature of early attempts to consider effects of magnetic field in atmospheres modelling, as well as a simplified but state-of-the-art technique of stellar model atmospheres calculation taking into account the magnetic absorption, were presented in our previous works \citep[][hereafter \paper1]{khan_paper1,paper1}.

The development of improved model atmospheres for CP stars is an extremely important task for understanding the nature of features of CP stars. Model atmospheres are used for a variety of stellar astrophysics problems in application to CP stars: fundamental parameters determination, abundance analysis, stellar magnetic field geometry research, detailed line profile study (including the full treatment of the Zeeman effect), and stellar surface properties reconstruction by the Doppler Imaging technique (including all Stokes parameters). An enormous part of this work is currently being performed using classic model atmospheres.

In general, the magnetic field affects the energy transport, hydrostatic equilibrium, diffusion processes and the formation of spectral lines. In \paper1\ we considered in detail the last effect, due to complex Zeeman splitting, assuming a simplified model of a magnetic field vector perpendicular to the line of sight and estimated its influence on model atmospheres without solving the polarized radiation equation.

This paper continues the investigation and aim started in the first paper of this series: to introduce a realistic calculation of the anomalous Zeeman effect in the classical 1-D models of stellar atmospheres and to investigate the resulting effects on the model structure, energy distribution and other general observables.
In this work we summarize the transfer equation for polarized radiation, describe the general technique of our computational procedure, and calculate the same grid of model atmospheres of A and B stars as in \paper1\, exploring a $\tef$ range relevant for CP stars. The paper is organized as follows. In Sect.\,\ref{techniques} we
describe our technique of model atmosphere calculation and numerical implementation of the transfer equation for polarized radiation in the line opacity calculation. Section\,\ref{results} presents results of numerical application and, Sect.\,\ref{discussion} summarizes our work.

\section{Calculation of magnetic model atmospheres}
\label{techniques}
The model atmospheres calculated in this paper were constructed using a modified version of the \llm\ code, originally developed by \citet{llmodels}. The \llm\ code is an LTE (local thermodynamical equilibrium) $\mbox{1-D}$ stellar model atmosphere code for early and intermediate type stars which is intended for as accurate a treatment as possible of the line opacity using a direct method for the line blanketing calculation. This approach allows us to take into account individual chemical compositions (not scaled to the solar) of stellar atmospheres, possibly inhomogeneous vertical distribution of abundances (as a result of diffusion processes which we do not consider here in a self-consistent way), and individual line broadening mechanisms such as anomalous Zeeman splitting of spectral lines that are required for proper modelling of atmospheres of CP stars.

Using other techniques (Opacity Distribution Function, Opacity Sampling) for modelling atmospheres of CP stars is very complicated or even impossible due to their basic limitations and statistical nature. These methods were developed several decades ago in order to reduce the computational time of the treatment of millions of spectral lines.

In contrast, the direct method implemented in the \llm\ code is free of any approximations so that it fully describes the dependence of line absorption coefficient upon frequencies and depths in a model atmosphere, it does not require pre-calculated opacity tables, it has no limits on number of frequency points or number of spectral lines, and it has very reasonable execution time on an ordinary PC.

The code is based on modified \atlas\ subroutines \citep{kurucz13} and on
the spectrum synthesis code described by \citet{tsymbal}. In the new version of the \llm\ code, the continuum
opacity sources and partition functions of iron-peak elements from {\sc ATLAS12} \citep{atlas12} are used.
Parts of the magnetic spectrum synthesis code \synthm\ \citep{khan_synthm} were incorporated into \llm\ to deal with transfer of polarized radiation.

\subsection{Transfer equation for polarized radiation}
In this subsection we summarize the basics of radiative transfer in the presence of the magnetic field, formulated in suitable way for our calculations, and point out some of its important details. The comprehensive description and math for polarized radiation can be found in several books \citep{rees_nrt1,rees_nrt2,stenflo,toro}.

The polarized beam can be fully described in term of Stokes formalism
\begin{equation}
\vec{I}=(I,Q,U,V)^\dag,
\end{equation}
where $\vec{I}$ is called Stokes vector.
The parameter $I$ represents the total intensity of the beam, $Q$ and $U$ the difference between the intensities of linearly polarized components along position angles ($0,\pi/2$) and ($\pi/4,3\pi/4$), and $V$ the intensity difference between right-handed and left-handed circular polarized components.

The Stokes formulation of the transfer equation on a scale of column mass $m$ (which is usually used in model atmospheres calculation) is
\begin{equation}
\mu\frac{{\rm d}\vec{I}}{{\rm d}m}=\vec{K}\vec{I}-\vec{j},
\label{pol_eq}
\end{equation}
where $\mu\equiv\cos(\theta)$ is the cosine of the angle $\theta$ between the light propagation direction and the perpendicular to the surface, $\vec{K}$ is the total propagation matrix for all contributing lines,
\begin{equation}
\vec{K}=(\kappa_{\rm c}+\sigma_{\rm c})\vec{1}+\sum_{\rm lines}\kappa_0^{\rm line}\vec{\Phi}_{\rm line}.
\label{propagation_matrix}
\end{equation}
If we assume unpolarized continuum radiation and consider coherent isotropic scattering, neglecting polarization effects, then
the total emission vector $\vec{j}\,$ is
\begin{equation}
\vec{j}=\kappa_{\rm c}\, S_{\!\rm c}\vec{1}_0+S_{\!\rm L}\vec{1}_0\sum_{\rm lines}\kappa_0^{\rm line}\vec{\Phi}_{\rm line}+\sigma_c{\rm J_\nu}\vec{1}_0.
\label{emission_vector}
\end{equation}
Here $\vec{1}$ is a ${4\times 4}$ identity matrix, $\vec{1}_0=(1,0,0,0)^\dag$, $\kappa_c$ and $S_{\!\rm c}$ are the continuum mass absorption coefficient and source function, $\kappa_0$ is the line center mass absorption coefficient for zero damping, zero magnetic field and corrected for stimulated emission, $\sigma_c$ is continuous scattering coefficient, and ${\rm J}_\nu=\oint I_\nu\,{\rm d}\omega/4\pi$ is mean intensity.

If we assume LTE then, ${S_{\!\rm c}=S_{\!\rm L}=B_\nu}$, the Planck function at the local electron temperature. In that case introducing the source vector
\begin{equation}
\vec{S}=(1-\alpha)B_\nu\vec{1}_0+\alpha\,{\rm J}_\nu\vec{1}_0,
\end{equation}
where $\alpha=\sigma_c/\kappa_I$, we can rewrite Eq.\,(\ref{pol_eq}) as
\begin{equation}
\mu\frac{{\rm d}\vec{I}}{{\rm d}m}=\vec{K}(\vec{I}-\vec{S}).
\label{pol_eq1}
\end{equation}
To compute the source vector $\vec{S}$ we perform $\Lambda$-iterations as described by \citet{kurucz_atlas9}.

\begin{figure}
\fifps{5cm}{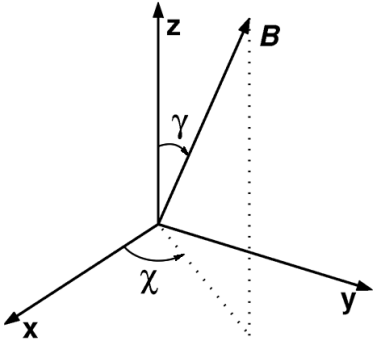}
\caption{Reference frame for representing the Stokes vector $\vec{I}$ and magnetic field vector $\vec{B}$. The $z$-axis shows the direction of the beam propagation (or line of sight). The magnetic vector has an inclination $\gamma$ to the propagation direction and azimuth $\chi$ measured counter-clockwise from the $x$-axis: $0\leq\gamma\leq\pi$\,,\,\,\,$0\leq\chi\leq 2\pi$.}
\label{reference1}
\end{figure}

Finally, the line propagation matrix $\vec{\Phi}$ is
\begin{equation}
\begin{array}{rcll}
\vec{\Phi}= & \left(
\begin{array}{llll}
\phi_I & \,\,\,\, \phi_Q & \,\,\,\, \phi_U & \,\,\,\, \phi_V \\
\phi_Q & \,\,\,\, \phi_I & \,\,\,\, \psi_V &         -\psi_U \\
\phi_U &         -\psi_V & \,\,\,\, \phi_I & \,\,\,\, \psi_Q \\
\phi_V & \,\,\,\, \psi_U &         -\psi_Q & \,\,\,\, \phi_I \\
\end{array}
\right)
\end{array},
\end{equation}
where
\begin{equation}
\begin{array}{ll}
\phi_I= & \frac{1}{2}\,(\phi_p\sin^2\gamma+\frac{1}{2}(\phi_r+\phi_b)(1+\cos^2\gamma)) \vspace{0.2cm},\\
\phi_Q= & \frac{1}{2}\,(\phi_p-\frac{1}{2}(\phi_r+\phi_b))\sin^2\gamma\,\cos2\chi \vspace{0.2cm},\\
\phi_U= & \frac{1}{2}\,(\phi_p-\frac{1}{2}(\phi_r+\phi_b))\sin^2\gamma\,\sin2\chi \vspace{0.2cm},\\
\phi_V= & \frac{1}{2}\,(\phi_r-\phi_b)\cos\gamma \vspace{0.2cm},\\
\psi_Q= & \frac{1}{2}\,(\psi_p-\frac{1}{2}(\psi_r+\psi_b))\sin^2\gamma\,\cos2\chi \vspace{0.2cm},\\
\psi_U= & \frac{1}{2}\,(\psi_p-\frac{1}{2}(\psi_r+\psi_b))\sin^2\gamma\,\sin2\chi \vspace{0.2cm},\\
\psi_V= & \frac{1}{2}\,(\psi_r-\psi_b)\cos\gamma.
\end{array}
\label{matrix_coefficients}
\end{equation}
The angles $\gamma$ and $\chi$ are defined in the cartesian reference frame $xyz$ as shown in Fig.~\ref{reference1}. The $\phi_{p,b,r}$ and $\psi_{p,b,r}$ are absorption and anomalous dispersion profiles that will be described below.

For a better understanding of the physical meaning of the $\vec{\Phi}$ matrix, it can be presented as sum of three matrices:
\begin{equation}
\vec{\Phi}=\phi_I\vec{1}+%
\left(%
\begin{array}{llll}
     0 & \phi_Q & \phi_U & \phi_V \\
\phi_Q &      0 &      0 &      0 \\
\phi_U &      0 &      0 &      0 \\
\phi_V &      0 &      0 &      0 \\
\end{array}
\right)+%
\left(%
\begin{array}{llll}
0 & \,\,\,\,\,\,      0 & \,\,\,\,      0 & \,\,\,\,      0 \\
0 & \,\,\,\,\,\,      0 & \,\,\,\, \psi_V &         -\psi_U \\
0 & \,\,        -\psi_V & \,\,\,\,      0 & \,\,\,\, \psi_Q \\
0 & \,\,\,\,\,\, \psi_U &         -\psi_Q & \,\,\,\,      0 \\
\end{array}
\right).
\end{equation}
The first, diagonal, matrix corresponds to absorption by the medium that results in equal changes of all Stokes parameters. The second, symmetric, matrix corresponds to dichroism, where some polarized components are extinguished more than others. The third, antisymmetric matrix corresponds to dispersion phenomena (or birefringence), where phase shifts change states of linear polarization among themselves (Faraday rotation) and states of linear polarization with states of circular polarization (Faraday pulsation). The sum of two first matrices represents absorption effects, the last matrix represents anomalous dispersion (or magneto-optical effects).

The choice of $x$ and $y$ axes in the reference frame $xyz$ (Fig.~\ref{reference1}) is free. In fact the intensity of the beam and degrees of linear $\sqrt{Q^2+U^2}/I$ and of circular $V/I$ polarization are independent of the choice of $x$ and $y$ axes. Specifically, the rotation of the reference frame by an angle $\alpha$ measured in same direction as $\chi$ modifies the Stokes vector $\vec{I}$ as
\begin{equation}
\vec{I}'=\vec{G}\vec{I},
\end{equation}
where
\begin{equation}
\vec{G}(\alpha)=\left(
\begin{array}{cccc}
1 & 0 & 0 & 0 \\
0 & \,\,\,\,\,\cos 2\alpha & \sin 2\alpha & 0 \\
0 & -\sin 2\alpha & \cos 2\alpha & 0 \\
0 & 0 & 0 & 1
\end{array}
\right).
\label{rotation_matrix}
\end{equation}
Hence neither the total beam intensity $I$ nor the $V$ parameter depend on the $xy$ choice, whereas $Q'$ and $U'$ are defined by simple linear transformations.

Now, let us return to the absorption $\phi_{p,b,r}$ and the anomalous dispersion $\psi_{p,b,r}$ profile definitions.
In the presence of a magnetic field an atomic level $k$ defined by the  quantum
numbers $J_k$, $L_k$, $S_k$ splits into the $2J_k+1$ states with the magnetic
quantum numbers $M_k=-J_k, \dots, +J_k$. The absolute value of the splitting is
defined by the field modulus $|\vec{B}|$ and by the Land\'e factor $g_k$,
which in the case of LS coupling can be calculated as
\begin{equation}
g_k=\frac{3}{2}+\frac{S_k(S_k+1)-L_k(L_k+1)}{2J_k(J_k+1)}.
\end{equation}
According to the selection rules the following transitions are allowed between
the split upper $u$ and lower $l$ levels
\begin{equation}
\Delta{M}=M_u-M_l=\left\{
\begin{array}{rcl}
+1&\equiv&b\\0&\equiv&p \\-1&\equiv&r
\end{array}.
\right.
\end{equation}

The wavelength shift of the component $i_j$ (where ${j=p,b,r}$, and
${i_{p,b,r}=1, \dots, N_{p,b,r}}$;\,\,$N_{p,b,r}$ is the number of components in
each group) relative to the laboratory line centre $\lambda_0$ is defined by
\begin{equation}
\Delta\lambda_{i_j}=\frac{e\lambda_0^2\,|\vec{B}|}{4\pi mc^2}\,(g_lM_l-g_uM_u)_{i_j}\,.
\label{lambda_shift}
\end{equation}

The relative strengths $S_{i_j}$ of the $\pi$ and $\sigma$ components
are given by \citet{sobelman} (see Table\,2 in \paper1) and are proportional to
\begin{equation}
\left(
\begin{array}{lcl}
\,\,\,\,J_u & 1 & J_l \\
-M_u & M_u-M_l & M_l
\end{array}
\right)^2,
\end{equation}
a $3\rm j$ - symbol.
The normalization of the relative strengths is performed to unity for each group of the Zeeman components
\begin{equation}
\sum\limits_{i=1}^{N_p}S_{i_p}=\sum\limits_{i=1}^{N_b}S_{i_b}=\sum\limits_{i=1}^{N_r}S_{i_r}=1.
\end{equation}
The absorption and anomalous dispersion profiles in absence of macroscopic velocity field are
\begin{equation}
\phi_j=\sum\limits_{i_j=1}^{N_j}S_{i_j}H(a,\,\upsilon-\upsilon_{i_j}) 
\end{equation}
and
\begin{equation}
\psi_j=2\sum\limits_{i_j=1}^{N_j}S_{i_j}F(a,\,\upsilon-\upsilon_{i_j}), 
\end{equation}
where $H(a,\upsilon)$ and $F(a,\upsilon)$ are the Voigt and Faraday-Voigt functions \citep[][p.\,65]{stenflo}
\begin{equation}
H(a,\upsilon)=\frac{a}{\pi}\int\limits_{-\infty}^{+\infty}\frac{{\rm e}^{-y^2}}{(\upsilon-y)^2+a^2}\,{\rm d}y
\end{equation}
and
\begin{equation}
F(a,\upsilon)=\frac{1}{2\pi}\int\limits_{-\infty}^{+\infty}\frac{(\upsilon-y){\rm e}^{-y^2}}{(\upsilon-y)^2+a^2}\,{\rm d}y.
\end{equation}
The parameters $a$, $\upsilon$ and $\upsilon_{i_j}$ are expressed in units of the Doppler width $\Delta\lambda_{\rm D}$
\begin{equation}
a=\Gamma\lambda_0^2/4\pi c\Delta\lambda_{\rm D},
\end{equation}
where $\Gamma$ is the total line damping parameter,
\begin{equation}
\upsilon=(\lambda-\lambda_0)/\Delta\lambda_{\rm D},
\end{equation}
where $\lambda$ is the current wavelength in the line, and
\begin{equation}
\upsilon_{i_j}=\Delta\lambda_{i_j}/\Delta\lambda_{\rm D}.
\end{equation}

\subsection{Solution of the transfer equation}
Usually, one of two powerful methods to solve polarized transfer equation are used: the integral Feautrier method and the differential DELO (Diagonal Element Lambda Operator) method \citep{ahh,rees,piskunov}, which have proved their capabilities for accurate solution of the transfer equation. However, the Feautrier method is only half as efficient as DELO \citep{rees, piskunov}. Considering this fact, and the requirement that for stellar atmosphere modelling in comparison to spectrum synthesis, one needs to know not only the emergent Stokes vector at the surface but the radiation field (flux and mean intensity) through out the atmosphere, that requires the solution of an additional differential equation (and extra computational time), the method of choice is DELO.

The DELO method is a one-way method that allows one to calculate inward and outward radiation separately. This method, as a differential method, does not require special properties of the propagation matrix: the presence of symmetry in directions and frequencies. It was developed by \citet{rees} who used the fact that the diagonal elements of matrix $\vec{K}$ are equal to
\begin{equation}
\kappa_I=\kappa_c+\sigma_c+\sum_{\rm lines}\kappa_0^{\rm line}\phi_I^{\rm line}.
\end{equation}
Defining a vertical optical depth as
\begin{equation}
{\rm d}\tau=\kappa_I\,{\rm d}m,
\end{equation}
a modified propagation matrix $\vec{K}'$ with zeros on its diagonal
\begin{equation}
\vec{K}'=\vec{K}/\kappa_I-\vec{1}
\end{equation}
and a modified vector
\begin{equation}
\vec{S}'=\vec{j}/k_I,
\end{equation}
which has the units of a source function, we can rewrite Eq.~(\ref{pol_eq}) in the form
\begin{equation}
\mu\,\frac{{\rm d}\vec{I}}{{\rm d}\tau}=\vec{I}-\vec{\xi},
\label{delo_eq}
\end{equation}
where
\begin{equation}
\vec{\xi}=\vec{S}'-\vec{K}'\vec{I}
\label{source_term}
\end{equation}
is the new source term (or effective source function).

Let us introduce a grid of depth points from the surface to the deepest layer ${\tau_i\,\,(i=1,...,N)}$ along the vertical direction. The formal integral solution of Eq.~(\ref{delo_eq}) can be written as relation between the Stokes vectors $\vec{I}_i$ and $\vec{I}_{i+1}$,
\begin{equation}
\vec{I}_i=\epsilon_i\vec{I}_{i+1}+\int\limits_{\tau_i}^{\tau_{i+1}}{\rm e}^{-(\tau-\tau_i)}\vec{\xi}(\tau)\,{\rm d}\tau,
\label{delo_integral}
\end{equation}
using a simple trapezoidal formula
\begin{equation}
\begin{array}{l}
\delta_i=\frac{1}{2}(\kappa_{I}^{\,i+1}+\kappa_{I}^{\,i})(m_{i+1}-m_i)/\mu \vspace{0.1cm}, \\
\epsilon_i=e^{-\delta_i}.
\end{array}
\end{equation}
In order to solve the integral (\ref{delo_integral}) analytically two approaches were developed. The first one, originally proposed by \citet{rees}, is based on linear interpolation of the source term $\vec{\xi}$ between grid points, while the second one relies on linear interpolation only for $\vec{K}'\vec{I}$ and parabolic approximation for $\vec{S}'$ \citep{socas}. The solution of Eq.~(\ref{delo_integral}) for these two cases can be written in common form
\begin{equation}
X_i\vec{I}_i=Y_i\vec{I}_{i+1}+Z_i,
\label{delo_solution}
\end{equation}
where
\begin{equation}
\begin{array}{l}
X_i=\vec{1}+(\alpha_i-\beta_i)\vec{K}'_i, \\
Y_i=\epsilon_i\vec{1}-\beta_i\vec{K}'_{i+1}
\end{array}
\end{equation}
and
\begin{equation}
\begin{array}{l}
\alpha_i=1-\epsilon_i, \vspace{0.1cm} \\
\beta_i=[1-(1+\delta_i)\epsilon_i]/\delta_i.
\end{array}
\end{equation}
For a linear approximation of the source term $\vec{S}'$
\begin{equation}
Z_i=(\alpha_i-\beta_i)\vec{S}'_i+\beta_i\vec{S}'_{i+1},
\end{equation}
and for a quadratic approximation
\begin{equation}
Z_i=a_i\vec{S}'_{i-1}+b_i\vec{S}'_{i}+c_i\vec{S}'_{i+1},
\end{equation}
where \citep[see][]{piskunov}
\begin{equation}
\begin{array}{l}
a_i=\displaystyle\frac{z-\delta_{i+1}y}{(\delta_i+\delta_{i+1})\delta_i}, \vspace{0.1cm}\\
b_i=\displaystyle\frac{(\delta_{i+1}+\delta_i)y-z}{\delta_i\delta_{i+1}}, \vspace{0.1cm}\\
c_i=x+\displaystyle\frac{z-(\delta_i+2\delta_{i+1})y}{\delta_{i+1}(\delta_i+\delta_{i+1})}, \vspace{0.1cm}\\
x=1-\epsilon_i, \vspace{0.1cm}\\
y=\delta_i-x, \vspace{0.1cm}\\
z=\delta_i^2-2y.
\end{array}
\end{equation}
We used the quadratic DELO method because it is significantly more accurate that linear DELO \citep{socas,piskunov}.

The recursive Eq.~(\ref{delo_solution}) is used to calculate the Stokes vectors of radiation directed outwards $\vec{I}^+$ or inwards $\vec{I}^-$ through the atmosphere with different boundary conditions. To calculate $\vec{I}^+$ we start at the lower boundary assuming unpolarized thermalized radiation $\vec{I}^+_N=B_{\nu,N}\vec{1}_0$, or using the improved asymptotic formula \citep{ahh}
\begin{equation}
\vec{I}^+_N=B_{\nu,N}\vec{1}_0+\mu\,\vec{K}_N^{-1}\vec{1}_0\left(\frac{{\rm d}\!B_\nu}{{\rm d}m}\right)_N,
\label{bound_n}
\end{equation}
and iterating outwards to the surface.
The calculation of the $\vec{I}^-$ starts with initial estimate
\begin{equation}
\vec{I}^-(0)=0,
\label{bound_1}
\end{equation}
assuming that there is no inward radiation at the surface, and iterates inwards to the atmosphere.

\subsection{Model convergence criteria}
\label{conveg_par}
To control convergence of the model calculations we use two criteria: the condition of radiative equilibrium and constancy of the total flux \citep{tsymbal_ll,llmodels}. Both criteria are checked on each iteration and for each atmospheric layer, and the iterative procedure continues as long as errors are more then 1\% and and temperature correction exceed 1K.

The condition of the radiative equilibrium (or total energy balance) expressing the absence of sources and sinks of energy may be easily derived from Eq.~(\ref{pol_eq1}) by integration over solid angle and frequency,
\begin{equation}
\frac{{\rm d}}{{\rm d}m}\int\limits_0^\infty\oint\vec{I}_\nu\cos\theta\,{\rm d}\omega\,{\rm d}\nu=
\int\limits_0^\infty\oint\vec{K_\nu}(\vec{I_\nu}-\vec{S_\nu})\,{\rm d}\omega\,{\rm d}\nu.
\label{rad_eq1}
\end{equation}
The left side of the first equation of this system is
\begin{equation}
\frac{{\rm d}}{{\rm d}m}\int\limits_0^\infty\oint I_\nu\cos\theta\,{\rm d}\omega\,{\rm d}\nu=\frac{{\rm d}}{{\rm d}m}F(m)=0,
\end{equation}
where $F(m)$ is the total flux at depth $m$ in an atmosphere.
Thus the right side of the first equation of the system (\ref{rad_eq1}) leads to formulation of radiative equilibrium as
\begin{equation}
\begin{array}{l}
\int\limits_0^\infty\oint(\kappa_I I+\kappa_Q Q+\kappa_U U+\kappa_V V)\,{\rm d}\omega\,{\rm d}\nu=\int\limits_0^\infty\oint\kappa_I\,S_{\!\nu}\,{\rm d}\omega\, {\rm d}\nu.
\end{array}
\label{rad_eq}
\end{equation}
Here we note that the integrand expression of Eq.~(\ref{rad_eq}) does not depend on reference frame rotation, which followes from Eqs.~(\ref{matrix_coefficients}) and (\ref{rotation_matrix}).

The flux conservation condition is formulated in the usual manner
\begin{equation}
\int\limits_0^\infty F_\nu\,{\rm d}\nu-\sigma T_{\rm eff}^4=0,
\end{equation}
where $F_\nu$ can be calculated as an intensity integral over solid angle ${\rm d}\omega=\sin\theta\,{\rm d}\theta\,{\rm d}\varphi=-{\rm d}\mu\,{\rm d}\varphi$ as follows
\begin{equation}
F_\nu=\int\limits_0^{2\pi}\int\limits_0^{+1}\mu\,I^+ {\rm d}\mu\,{\rm d}\varphi-\int\limits_0^{2\pi}\int\limits_0^{+1}\mu\,I^- {\rm d}\mu\,{\rm d}\varphi,
\label{flux}
\end{equation}
where $I^+$ and $I^-$ are the first elements of corresponding Stokes vectors $\vec{I^+}$ and $\vec{I^-}$ which are calculated in Eq.~(\ref{delo_solution}) with boundary conditions (\ref{bound_n}) and (\ref{bound_1}), and angles $\theta$ and $\varphi$ are represented in Fig.~\ref{reference2}. Here we neglect convective flux because the strong global magnetic field may be supposed to prevent turbulent motions in the atmospheres of Ap and Bp stars.

\subsection{Optical depth scale}
In model atmosphere calculations with radiative transfer of polarized radiation we can not use Rosseland depths $\tau_{R}$ as an optical depth scale because the extinction is not represented by a scalar but by a matrix. Moreover, this matrix depends on light propagation direction in the atmosphere, Eq.~(\ref{matrix_coefficients}).

However, keeping in mind our assumption (\ref{emission_vector}) that the continuum is unpolarized it is convenient to introduce a continuum optical depth scale $\tau_c$, instead of using $\tau_{R}$, by
\begin{equation}
{\rm d}\tau_c=(\kappa_c+\sigma_c)\,{\rm d}m.
\end{equation}
We have used the monochromatic optical depth $\tau_c$ calculated at $5000$\,\AA\, denoted here as $\taucont$, as a scale on which atmosphere model physical quantities are specified. The wavelength $5000$\,\AA\ is usually used as a monochromatic optical depth in analysis of stellar atmospheres, and its choice is determined by the fact that the continuum opacity does not vary rapidly at this frequency and it is located around the maximum of the energy flow. Besides, the same depth scale is used in the \llm\ code for solution of hydrostatic equation \citep{llmodels}.

The model atmospheres calculations are carried out on a fixed $\taucont$ grid using equal-space $\log\taucont$ scale subdivided into layers.

\subsection{Coordinate systems}
\begin{figure}
\fifps{5cm}{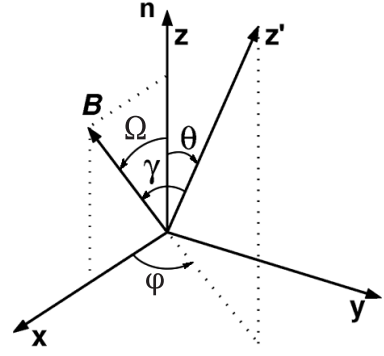}
\caption{Reference frame of plane-parallel atmosphere. The $z$-axis is parallel to the unit vector $\vec{n}$ and the $xy$ plane coincides with the plane of an atmosphere. The $z'$-axis (propagation direction of light) is the same as $z$-axis in Fig.~\ref{reference1} and has an inclination $\theta$ to the unit vector $\vec{n}$ and azimuth $\varphi$ measured counterclockwise from the $x$-axis: ${0\leq\theta\leq\pi}$\,,\,\,\,${0\leq\varphi\leq 2\pi}$. The $\gamma$ angle has the same meaning as in Fig.~\ref{reference1}. The magnetic field vector $\vec{B}$ is in the $xz$ plane and has an inclination $\Omega$ to the unit vector: $0\leq\Omega\leq\pi$.}
\label{reference2}
\end{figure}

For a proper model atmosphere calculation in the presence of a magnetic field we need complete information about the orientation of the magnetic field vector $\vec{B}$ and the light propagation direction in the model reference frame. In Fig.~\ref{reference2} two cartesian references frames are represented. One of them $(xyz)$ is related to the plane-parallel atmosphere, another, local reference frame $(x'y'z')$ depends on the light propagation direction and the $z'$-axis coincides with this direction. The $x'$ and $y'$ axes are not shown in figure because their choice does not modify the main equations used in this paper (Sect.~\ref{conveg_par}). Thus, three angles: $\theta$ and $\varphi$ (inclination of the propagation direction to the unit vector $\vec{n}$, and its azimuth), and $\gamma$ (inclination of magnetic vector $\vec{B}$ to the propagation direction) in the model reference frame $xyz$ fully represent the necessary orientation related information.

The direction of the magnetic field vector in the model reference frame $xyz$ is set by the inclination angle $\Omega$. It is clear that an additional azimuthal angle is not necessary because of the integral nature of the flux and radiative equilibrium conditions, Eqs.~(\ref{rad_eq}) and (\ref{flux}). That is why it is convenient to introduce the model reference frame $xyz$ so that the $x$-axis is in the $(\vec{B},\vec{n})$ plane. In this case the angle $\gamma$ is defined by formula
\begin{equation}
\cos\gamma=\cos\theta\cos\Omega+\sin\theta\sin\Omega\cos\varphi.
\end{equation}

Consequently, one preset angle $\Omega$ and the spherical coordinate system with angles $\theta$ and $\varphi$ completely describe the orientation of the magnetic field vector for computing models.

\section{Results of numerical application}
\label{results}
In order to test model atmospheres including polarized radiation transfer we calculated the same set of model atmospheres for A and B stars with the same modelling parameters as in \paper1. The main aim of this numerical application is to analyse effects of introduction of the polarized transfer equation in model atmosphere computation, because we suppose that the treatment of Zeeman splitting used in our previous work overestimates the magnetic intensification.

In \paper1 we treated individual Zeeman components of the anomalous splitting pattern as independent lines by modification of the original lines list and $gf$ values for these lines. Splitting pattern incorporated into the line list supposes that the orientation of the magnetic field vector is characterized by the same angle $\gamma$ for every light propagation direction. That means that we neglected anisotropy, which is naturally arises in magnetic atmospheres and is determined by the orientation of the magnetic field vector. In \paper1 we assumed the angle $\gamma$ is always equal to $\pi/2$ to minimize effects of the polarized radiative transfer. That allowed us to use the transfer equation for non-polarized radiation and simplified the task.

In other words, such an approach supposes that the magnetic field orientation is assigned not with respect to the plane of atmosphere (or observer), but to the light propagation direction. That is, the propagation matrix (\ref{propagation_matrix}) does not depend on angle $\theta$, and angle $\gamma$ is fixed (Fig.~\ref{reference2}). On the one hand this is an unrealistic configuration of the magnetic field; on the other hand, for spherical stars, the angle dependent Eq.~(\ref{pol_eq}) calculated in the plane-parallel approximation in an isotropic medium produces variation of the intensity across visible stellar disk that is the same as the local angular variation. This applied magnetic field configuration was called ``horizontal" in \paper1.

Using the techniques detailed in Sect.~\ref{techniques} and taking into account considerations outlined above to correspond to the previous investigation, we have calculated model atmospheres with the effective temperature
${\tef=8000}$\,K, 11\,000\,K, 15\,000\,K, surface gravity $\log g=4.0$, metallicity $[M/H]=0.0$, +0.5, +1.0 (for non-solar metallicities the He abundance was decreased to normalize the sum of all abundances) and magnetic field strength 0, 1, 5, 10, 20 and 40\,kG, which covers a substantial part of the stellar parameter space occupied by the magnetic Ap and Bp stars.

The model atmosphere calculations were carried out on the continuum optical
depth grid spanning from $+2$ to $-6.875$ in $\log\taucont$ and subdivided into
72 layers. Convection was neglected as we discussed above. We also adopted zero microturbulent velocity in all calculations, because of direct inclusion of the magnetic intensification in the modelling of spectral lines and because the same approach was used in \paper1.

At the start of the model computation for each of the $\tef$, $[M/H]$
pairs studied, we have used a standard \atlas\ \citep{kurucz13} model atmosphere for the
spectral line preselection procedure and as an initial guess of the model structure. The
line selection threshold was set to 1\% as previously.

We used the iterative approach for the calculation of the magnetic model atmosphere grid presented here as described in \paper1. Each new model is calculated using a converged model with a weaker field strength (i.e. the 0\,kG model is used as an initial guess for the 1\,kG model, then the 5\,kG model is calculated starting from the 1\,kG model, etc.).

The following wavelength ranges for the spectrum synthesis were used: from
500\,\AA\ to 50\,000\,\AA\ for ${\tef=8000}$\,K, from 500\,\AA\ to
30\,000\,\AA\ for ${\tef=11\,000}$\,K and from 100\,\AA\ to 30\,000\,\AA\ for
${\tef=15\,000}$\,K. For all models we used a wavelength step of 0.1\,\AA,
which resulted in the total number of frequency points in the range between
295\,000 and 495\,000. It was shown \citep{llmodels} that a wavelength step 0.1\,\AA\ is quite sufficient for model atmosphere calculations with the fundamental parameters used in this paper.

In the following sections we describe preparation of the line lists, and we study the effects of the magnetic line blanketing on the common photometric and spectroscopic observables: the spectral energy distribution, photometric colors and profiles of the hydrogen Balmer lines. We do not consider model structure (i.e. temperature and pressure distribution) because we have used an optical depth scale different from that which was used in \paper1. We generally believe that all main trends concerning changes of model structure due to backwarming outlined in \paper1 apply to this study too. Thus, those who are interested in model structure changes due to Zeeman effect should be referred to \paper1 (Sect.~3.1).

\subsection{Line lists and preselection procedure}
The technique for line list preparation is described in \paper1, and we used the same one. The magnetic line blanketing was taken into account for all spectral lines except the hydrogen lines according to the individual anomalous Zeeman splitting pattern.
The initial line list was extracted from VALD \citep{vald} including lines originating from predicted levels. The total number of spectral lines
was more than 22.5 million for the spectral range between 50 and 100\,000\,\AA. For $0.55$\% of these lines information about Land\'e factors was absent. Using the term designation
represented in the VALD and assuming the LS coupling approximation, we computed Land\'e factors for the lines of light elements, from He to Sc. That allowed us to reduce the number of lines with unknown Land\'e factors (or without a proper term designation) to $0.26$\%, for which we assumed a classical Zeeman triplet splitting pattern with effective Land\'e factor ${g_{\rm eff}=1.2}$. The resulting line list was converted to special binary format file accepted by the \llm\ code.

The resulting line list was used for the preselection procedure in the \llm\ code using the selection threshold 1\% that means that the code
selects spectral lines for which ${\ell_\nu/\alpha_\nu\geq 1}$\%, where $\ell_\nu$ and $\alpha_\nu$ are the continuum and line absorption coefficients at the frequency $\nu$. During the preselection procedure we have used a standard \atlas\ \citep{kurucz13} model atmospheres for each of the $\tef$, $[M/H]$ pairs studied, assuming a non-magnetic case.

Preselection allowed us to decrease the number of spectral lines
involved in the line blanketing calculation to about 300\,000\,--\,800\,000,
depending on the model atmosphere parameters. The number of preselected lines is presented in Table~\ref{number_lines}.

\begin{table}
\centering
\caption{Description of the spectral line lists used in the line blanketing
calculations for different effective temperature $\tef$ and metallicity
$[M/H]\equiv\log(N_{\rm metals}/N_{\rm H}) -\log(N_{\rm metals}/N_{\rm H})_\odot$.
$N_{\rm lines}$ is the total number of spectral lines retained by the
preselection procedure and included in the line blanketing calculation.}

\begin{tabular}{lllllll}
\hline\hline
$\tef, \rm K$ & $[M/H]$ & $N_{\rm lines}$ \\
\hline
   8000   & 0.0 & 352\,389 \\
          & 0.5 & 513\,034 \\
          & 1.0 & 744\,190 \\
\hline
11\,000   & 0.0 & 305\,249 \\
          & 0.5 & 442\,366 \\
          & 1.0 & 652\,839 \\
\hline
15\,000   & 0.0 & 361\,256 \\
          & 0.5 & 533\,787 \\
          & 1.0 & 798\,154 \\
\hline
\end{tabular}
\label{number_lines}
\end{table}

\subsection{Energy distribution}
\label{energy}
\begin{figure*}
\filps{\hsize}{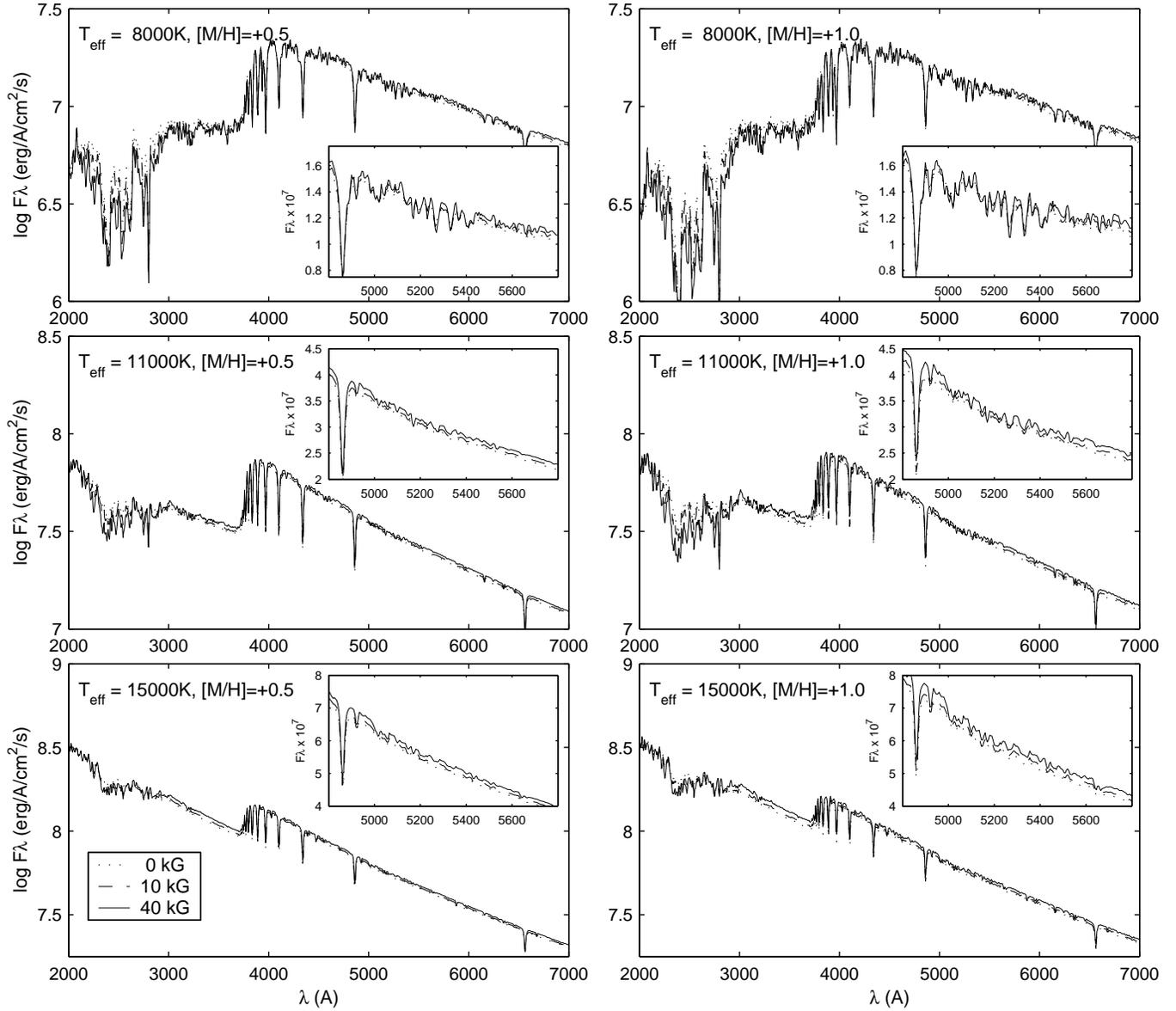}
\caption{Synthetic energy distributions from UV to near IR region for effective temperatures ${\tef=8000}$\,K, 11\,000\,K, 15\,000\,K and metallicities $[{M/H]=+0.5}$, ${[M/H]=+1.0}$. To enable visual comparison of the
different curves the original \llm\ stellar fluxes were convolved with a Gaussian profile with ${{\rm FWHM}=15}$\,\AA. The insets show energy distributions in the 5200\,\AA\ region.}
\label{fluxes}
\end{figure*}

The emergent spectral energy distributions $F_\nu$ (see Eq.~(\ref{flux})) for the three values of effective temperature and for metallicity ${[M/H]=+0.5}$ and ${[M/H]=+1.0}$ are presented in Fig.~\ref{fluxes}. The theoretical energy distributions for magnetic stars are
compared with the calculation for the reference non-magnetic models.

We found the same changes to the energy distributions compared to non-magnetic models as in \paper1. The first is
a flux deficiency in the ultraviolet spectral region coupled with a flux excess in the visual. The magnitude of the UV deficiency is a moderate function of the effective temperature and clearly increases with the magnetic field strength and metallicity.

The second related effect is that the presence of a magnetic field changes the stellar flux distribution in opposite directions in the visual and UV regions. At short wavelengths the magnetic star appears to be cooler in comparison to a non-magnetic object with the same fundamental parameters, whereas in the visual magnetic star mimics a hotter
normal star. The ``null wavelength'' where flux is unchanged progressively shifts to bluer wavelengths as the stellar effective temperature increases. The results of our modelling suggest that the ``null wavelength'' falls approximately at 3800\,\AA\ for ${\tef=8000}$\,K, at 2900\,\AA\ for ${\tef=11\,000}$\,K and at 2600\,\AA\ for ${\tef=15\,000}$\,K, and depends slightly on metallicity and field strength. The same ``null wavelengths'' were found in \paper1.

We found that the model atmospheres with magnetic line blanketing produce fluxes
that have a maximum deficiency in the UV region of about 0.1--0.2\,mag for ${B=10}$\,kG and
about 0.2--0.3\,mag for ${B=40}$\,kG, which is about half the effect found in \paper1.

The third effect is that the theoretical flux distribution of a magnetic star exhibits a depression in the 5200\,\AA\ region. This well-known depression is of interest because it is frequently observed in the spectra of peculiar stars and is used \citep{maitzen} to detect magnetic stars photometrically. In our theoretical calculations the 5200\,\AA\ depression is prominent at lower $\tef$ but becomes rather small for hotter models. The magnitude of the depression increases with the magnetic field intensity and metal content of the stellar atmosphere.

\subsection{Colors}
\label{color_sec}

\begin{figure*}[t]
\filps{\hsize}{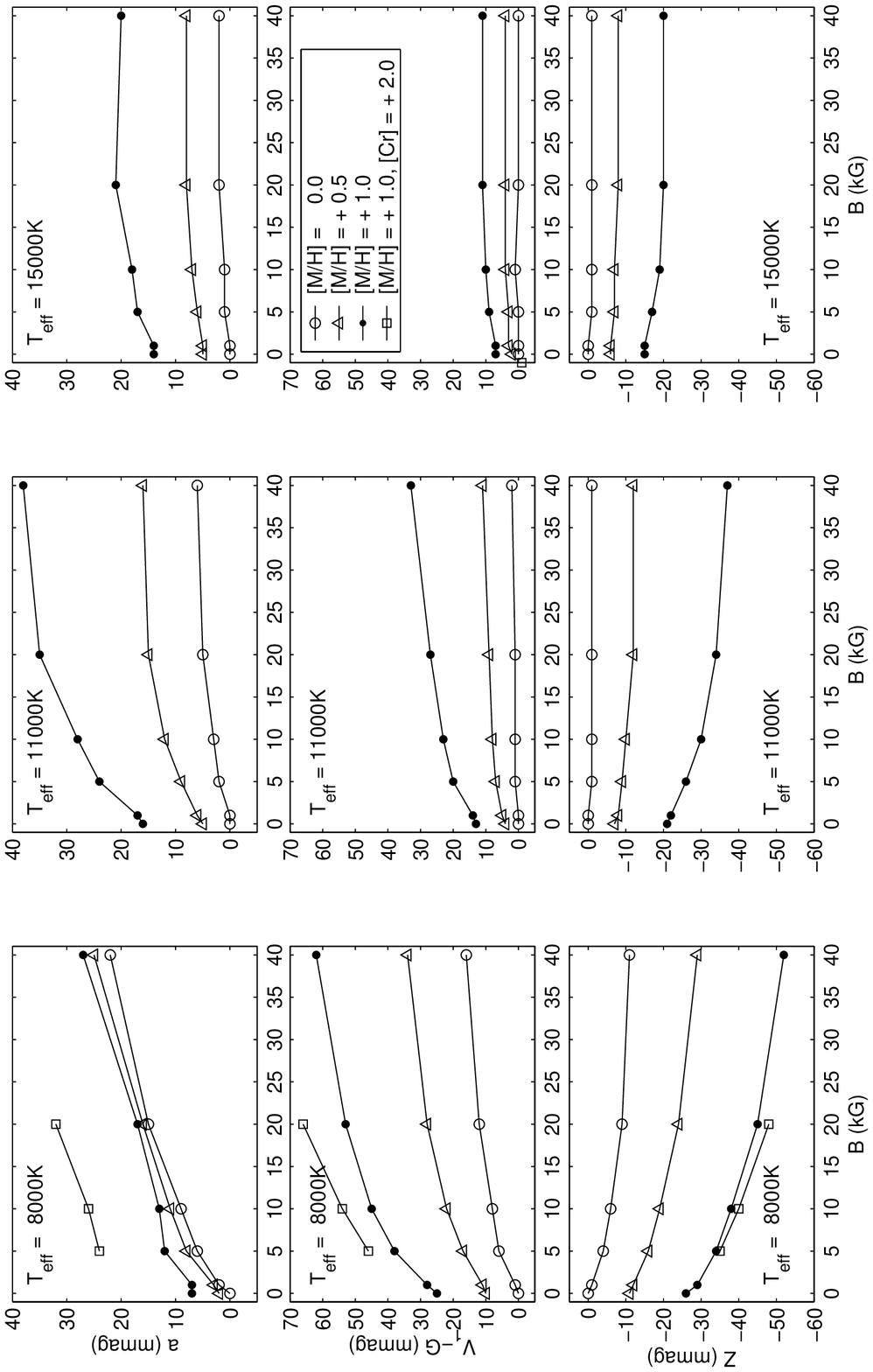}
\caption{The magnitude of the $a$, ${V_1-G}$ and $Z$ photometric indices as a function of the
magnetic field strength for ${\tef=8000}$, 11\,000 and 15\,000~K. Different curves
show calculations with metal abundances ${[M/H]=0.0}$ (open circles),
$[M/H]=+0.5$ (triangles) and $[M/H]=+1.0$ (filled circles). Open squares illustrate the effect of increasing Cr overabundance to ${[{\rm Cr}]=+2.0}$ in the ${\tef=8000}$\,K, ${[M/H]=+1.0}$ models. All curves are shifted so that for ${[M/H]=0.0}$ and zero magnetic field in each frame the $a$, ${V_1-G}$ and $Z$ indices equal zero.}
\label{peculiar}
\end{figure*}

We studied the influence of magnetic line blanketing on the photometric
colors in the Str\"omgren $uvby{\rm H}\beta$, Geneva and $\Delta a$ systems.

All colors were calculated using modified computer codes by \citet{kurucz13},
which take into account transmission curves of individual photometric filters,
mirror reflectivity and a photomultiplier response function. Our synthetic colors are computed from the energy
distributions sampled every 0.1\,\AA, so integration errors are expected to be negligible.

Having introduced a more accurate treatment of Zeeman splitting on line blanketing, it is particulary interesting to study specific photometric indices whose aim is to identify CP stars. Three such indices have been identified. The peculiar parameter $\Delta a$ was introduced by \citet{maitzen} to measure the strength of the 5200\,\AA\ depression and is based on narrow-band filters. The ${\Delta(V_1-G)}$ parameter of the Geneva medium-band system can be used as a peculiarity parameter for Ap stars \citep{hauck}. The $Z$ parameter, introduced by \citet{cramer1}, is linear combination of the Geneva photometric indices and was suggested as an indicator of chemical peculiarity. All of these indices are sensitive to the depression around 5200\,\AA\ and to the strength of the surface magnetic field.

The $\Delta a$ system was specifically designed for $5200$\,\AA\ depression measurement \citep{maitzen}, while the central wavelengths of the $V_1$ and $G$ filters of the ${\Delta(V_1-G)}$ system are located approximately at 5300\,\AA\ and 5800\,\AA\ respectively, so the region of sensitivity of the ${\Delta(V_1-G)}$ system is shifted to the red in comparison to $\Delta a$ system. The $Z$ index is mainly sensitive at $V_1$ filter \citep{cramer2}, so it is slightly shifted to the red, too.

Here we compare our results with those derived in \paper1, relying on fact that the behaviour of the photometric indices is closely related to the flux redistribution between the visual and UV regions and the presence of the flux depression at 5200\,\AA.

In order to save space for detailed illustrations of peculiarity indices we do not show general graphs for Str\"omgren $uvby{\rm H}\beta$ and the Geneva photometric systems, because their are not particularly interesting. However, we note some common features for all photometric indices. Relations between their changes and the intensity of magnetic field depends strongly on the stellar effective temperature. For
low $\tef$ photometric changes are very pronounced, whereas for hotter magnetic
stars modification of the photometric observables is fairly small (except $c_1$). All photometric indices become more sensitive to the magnetic effects with increasing metallicity. These main outcomes coincide with results obtained in \paper1, except that the sensitivity of all non-peculiar photometric indices is about 30--40\% less than those presented in the earlier paper.

The $c_1$ of the Str\"omgren system measures the Balmer discontinuity, which is why the shift of the ``null wavelength'', which leads to changes of the Balmer jump amplitude, affects the $c_1$ index. The relation between $c_1$ and $B$ demonstrates the same, almost linear, behaviour as in \paper1 but the overall effect is approximately half as large, which reflects weaker flux redistribution (see Sect.~\ref{energy}).

The results of our calculations of the synthetic peculiar $a$, ${V_1-G}$ and $Z$ photometric parameters are summarized in Fig.~\ref{peculiar} and show their sensitivity to the metal abundance and magnetic field strength. Similarly to the effects observed for other photometric indicators, these peculiar parameters are most influenced by magnetic field at lower $\tef$. In comparison to the results of \paper1 the sensitivity of these indices to the magnetic field strength appears to be between 10--50\% less as we decrease $\tef$ from 15\,000\,K to 8000\,K.

\begin{figure}
\filps{\hsize}{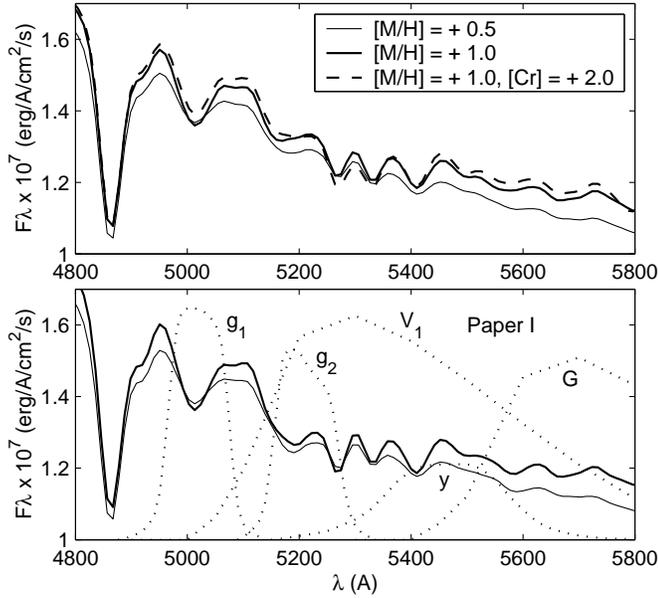}
\caption{Energy distribution around 5200\,\AA\ for effective temperature ${\tef=8000}$\,K, $B=20$\,kG and two values of metallicity ${[M/H]=+0.5}$ and ${[M/H]=+1.0}$; for second metallicity the behaviour for overabundant ${[{\rm Cr}]=+2.0}$ is also shown. The top frame represents the comparison of fluxes from present study, the bottom frame shows the respective data from \paper1. The fluxes are convolved with a Gaussian profile assuming resolution ${R=150}$. The curves of photometric filters (dash-dotted lines) are presented for comparison. The $g_1$, $g_2$ and $y$ are profiles of filters of the $\Delta a$ system, and $V_1$ and $G$ of the Geneva system.}
\label{color_flux}
\end{figure}

The $\Delta a$ system is most sensitive to the magnetic field strength, except for model atmospheres with ${\tef=8000}$\,K where the ${\Delta(V_1-G)}$ demonstrates the best sensitivity. Furthermore, it seems that the $\Delta a$ system is not very sensitive to metallicity (although it does increase with increasing $B$) for low $\tef$. While for non-magnetic case the small difference between $a$ indices for metallicities other then solar is normal feature \citep[][Fig.~4]{kupka1}, the small variation of $a$ with $[M/H]$ for strong magnetic field is not expected (in \paper1 we found strong dependence of the $a$-index on metallicity for all $\tef$). We suppose that this phenomena appears due to some saturation effect of enhanced line blanketing in the narrow region around 5200\,\AA, which is mainly dominated by the \ion{Fe}{I} and low excitation \ion{Fe}{II} lines for low effective temperature. In contrast to the narrow-band $\Delta a$-system the sensitivity of medium-band parameters ${V_1-G}$ and $Z$ does not suffer much from the specific properties of spectral feature in the narrow interval around $\lambda\,5200$\,\AA. This state is well illustrated in flux units (Fig.~\ref{color_flux}). One can see that flux depression in the region longward of 5200\,\AA\ is clearly affected by additional abundance and magnetic intensification, while the narrow region centered at 5200\,\AA\ is not altered much.

To test the influence of an anomalous concentration of an individual species
on the photometric peculiarity parameters, and especially the $\Delta a$ value for low effective temperature, we increased the Cr overabundance by a factor 100 relative to the sun and calculated additional model atmospheres.
As illustrated in Fig.~\ref{peculiar} and Fig.~\ref{color_flux} this results in 0.012--0.015~mag growth of the $\Delta a$ index for
${\tef=8000}$\,K, ${[M/H]=+1.0}$ model and field strength 5--20~kG. For ${\Delta(V_1-G)}$ the increase is smaller and amounts to 8--13~mmag while, the $Z$-index is almost unaffected (1--3~mmag). This behaviour indicates that observable anomalous $\Delta a$ values for low effective temperature may arise not only from the influence of the magnetic field but also from specific anomalous abundances as well. Nevertheless, even with overabundant Cr the $\Delta a$ system demonstrates the least sensitivity to the magnetic field strength, with only 8~mmag change from 5 to 20\,kG, while the value for $Z$ is 13~mmag and 20~mmag for ${\Delta(V_1-G)}$.

Finally, none of the photometric peculiarity indicators (except $a$ for ${\tef=8000}$\,K) show a linear trend over the whole range of the considered magnetic field strength. Saturation effects for their changes with the magnetic field intensity appear for ${B\gtrsim 10}$\,kG. The $\Delta a$ index for low effective temperature (${\tef=8000}$\,K) demonstrates almost linear dependency on $B$ from 5 to 40\,kG.

\subsection{Hydrogen line profiles}

\begin{figure*}[t*]
\filps{\hsize}{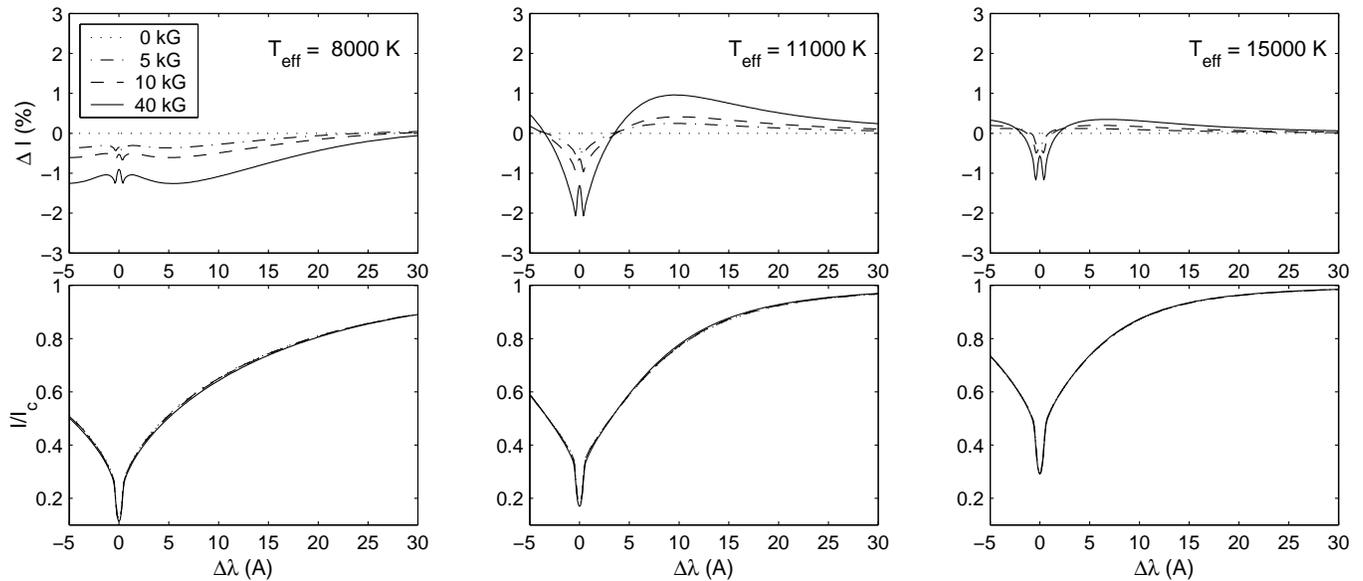}
\caption{Comparison between the synthetic H$\beta$ profiles computed for
${[M/H]=+1.0}$ and different values of the magnetic field strength and $\tef$.
The lower panels show normalized profiles, whereas the upper plots illustrate
the difference between H$\beta$ calculated for the models with substantial
magnetic line blanketing and the reference non-magnetic model.}
\label{hydrogen}
\end{figure*}

Profiles of the H$\alpha$, H$\beta$ and H$\gamma$ hydrogen Balmer lines were calculated
using the {\sc Synth} spectrum synthesis program \citep{piskunov_synth}.
The comparison between the magnetic
and non-magnetic profiles of the H$\beta$ line calculated for models with ${[M/H]=+1.0}$ metallicity is presented in Fig.~\ref{hydrogen}.

We found that the introduction of the polarized radiative transfer to the magnetic line blanketing calculation do not have a very strong influence on the Balmer line profiles. For proper further discussions we note that hydrogen line profiles calculated in this work for ${B=0}$\,kG are almost the same as those calculated in \paper1 (within $<1$\% error bar).

For all Balmer lines considered, the maximum change relative to the non-magnetic model does not exceed 1\% for ${B\le10}$~kG, which is true of the surface field strength for the great majority of magnetic CP stars, while the same quantity presented in \paper1\ (Sect.~3.4) is less then 2\%.

For metallicity ${[M/H]=0.0}$ the maximum change among all Balmer profiles relative to the non-magnetic models reaches 0.5\% of the continuum level for ${B\leq10}$\,kG and 1.0\% for ${B=40}$\,kG.

For the overabundant composition (${[M/H]=+1.0}$) the difference between hydrogen profiles for magnetic and non-magnetic models may be as large as 1.0\% for ${B\leq10}$\,kG and 1.8--2.5\% for ${B=40}$\,kG.

For the H$\beta$ line the maximum change amounts to about 2\% of the continuum level for ${B=40}$\,kG. The maximum deviation of 2.5\% found in the whole grid of models occurs for H$\alpha$ (${T_{\rm eff}=11\,000}$\,K, ${[M/H]=+1.0}$, and $B=40$\,kG), which exhibited the most sensitivity to magnetic field effects in Paper~I, too.

\subsection{Stellar atmospheric parameters and bolometric correction}
The modification of the energy distribution due to magnetic line blanketing may change photometric determination
of the stellar atmospheric parameters. To verify its possible influence on
$\tef$ and $\log g$ determination we applied the {\sc TempLogg} code \citep{templogg} which implements the calibration by \citet{napiwotski} to synthetic $uvby{\rm H}\beta$ colors (Sect.~\ref{color_sec}).

As in \paper1 we examined the error introduced by the magnetic field on the determination of the stellar atmosphere parameters. For each set of the effective temperature and metallicity we calculate differences between the values of $\tef$ and $\log g$ for non-magnetic and for the $B=40$\,kG models, then the maximum deviations of $\tef$ and $\log g$ among different metallicities are chosen. We found that for ${\tef=8000}$\,K the maximum error for $\tef$ and $\log g$ introduced by 40\,kG magnetic field is $-128$\,K and $-0.14$\,dex respectively. For the ${\tef=11\,000}$\,K models the temperature discrepancy is $+$81\,K and for $\log g$ is $-0.29$\,dex. For the hottest models (${\tef=15\,000}$\,K) the $\tef$ difference is $+502$~K, and for $\log g$ is $-0.12$\,dex.

At the same time the overabundant models demonstrate in most cases higher values of the effective temperature and $\log g$, but the differences in comparison to the solar composition models do not exceed 104\,K and 0.24\,dex for ${\tef=8000}$\,K, 106\,K and 0.13\,dex for ${\tef=11\,000}$\,K, and 538\,K and 0.15\,dex for ${\tef=15\,000}$\,K.

Finally, the differences between maximal and minimal values of the $\tef$ and $\log g$ for all values of the magnetic field strength and metallicity for ${\tef=8000}$\,K are 152\,K and 0.32\,dex, respectively, for ${\tef=11\,000}$\,K are 151\,K and 0.29\,dex, and for ${\tef=15\,000}$\,K are 868\,K and 0.16\,dex. The same values but for ${B\le10}$\,kG are smaller and equal to 104\,K and 0.25\,dex for ${\tef=8000}$\,K, 87\,K and 0.09\,dex for ${\tef=11\,000}$\,K, and 513\,K and 0.15\,dex for ${\tef=15\,000}$\,K.

We confirmed as in \paper1\ that even in the case of extreme magnetic field strengths, the bias in determination of the fundamental parameters using Str\"omgren photometry does not result in a systematic error in $\tef$ and $\log g$ beyond the usual error bars assigned to the photometrically determined stellar parameters.

We have not conducted the same estimate of possible influence on $\tef$ and $\log
g$ using Geneva photometric colors because in \paper1\ was shown that this system is less robust for determination of the atmospheric parameters of magnetic stars compared to the calibration in terms of the $uvby{\rm H}\beta$ colors.

Due to redistribution of the stellar flux from UV to visual, magnetic CP
stars appear brighter in $V$ compared to normal stars with the same fundamental
parameters. Using the theoretical flux distributions we
studied the effect of magnetic field and metallicity on the bolometric correction
(BC). For all stellar models this parameter changes relative to the non-magnetic case by
up to 0.029--0.053~mag for ${B=40}$\,kG and by no more than 0.018~mag for ${B\le10}$\,kG.
However, in the absence of the field, an increase of metallicity changes BC. For instance, for metallicity ${[M/H]=+1.0}$ the change is 0.084--0.125\,mag.

Generally, there no substantial differences in determination of atmospheric parameters and bolometric correction in comparison to the results obtained in \paper1.

\section{Conclusions}
\label{discussion}

This paper continues our study of magnetic line blanketing in stellar atmospheres and its relation to the observable quantities of magnetic CP stars. We presented a technique of model atmosphere calculation, introducing the transfer equation for polarized radiation. We have used a modified version of the model atmosphere code \llm, which implements a direct treatment of the line opacity.

Using this new technique, we computed a grid of model atmospheres of peculiar A and B stars which are similar to those computed in \paper1 in order to analyse effects of the introduction of the polarized transfer equation. We have used this grid to analyze the behaviour of observed characteristics for different magnetic fields, abundances and effective temperature.  To insure accurate modelling, we have used up-to-date atomic line data from the VALD database. This approach represents the most accurate treatment of the Zeeman effect in model atmosphere calculation used until now.

We found that the model atmospheres with magnetic line blanketing produce fluxes
that are deficient in the UV region, and the presence of a magnetic
field leads to flux redistribution from the UV to visual region.
This property of the theoretical models is in agreement with observations of CP
stars and with results presented in \paper1, but the overall effect is approximately twice smaller compared to the results of \paper1.

We also investigate the most prominent feature of the CP stars in the visual region, the flux depression centered at 5200\,\AA. Our numerical results show that the magnitude of the feature near 5200\,\AA\ depends strongly on magnetic field strength, $\tef$ and metallicity. The depression grows with increasing magnetic field strength and metallicity. However, for higher effective temperatures the magnitude of the depression for overabundant models becomes smaller, while the sensitivity to the magnetic field strength appears almost negligible. This outcome coincides with the conclusion found in the previous study.

In order to study the flux depression around 5200\,\AA\ more carefully we examined the influence of the magnetic field on the photometry observables, in particular, photometric peculiarity parameters $\Delta a$, $Z$ and ${\Delta(V_1-G)}$, that are sensitive in this spectral region. We found the same trends as in \paper1, i.e. changes of peculiar parameters due to the influence of the magnetic field are noticeable for low effective temperatures, whereas
for hotter stars sensitivity to the magnetic field is reduced considerably, and the relation between $B$ and the peculiarity index depends on metallicity.

The magnetic modification of these indices is less by 30--50\% (from higher to lower effective temperature) compared to those presented in \paper1. The $\Delta a$ system is still most sensitive to the magnetic field (the maximum change is 22\,mmag for the range of $B$ from 0 to 40\,kG and ${[M/H]=+1.0}$) for higher ${\tef=11\,000}$\,K or 15\,000\,K while for ${\tef=8000}$\,K the ${\Delta(V_1-G)}$ system is preferable (the maximum change is 38~mmag vs. 20~mmag for $\Delta a$ for field strength running from 0 to 40\,kG and ${[M/H]=+1.0}$). The $\Delta a$ system seems is not able clearly distinguish CP stars with metallicity other then solar for low effective temperature. Moreover, for low effective temperature (${\tef=8000}$\,K) and ${B=10}$\,kG (which is typical for the most strongly magnetic CP stars) it exhibits overall changes of only 9--13\,mmag, while ${\Delta(V_1-G)}$ and $Z$ for show overall changes of about 8--45\,mmag.

Our numerical tests show that Cr appears to be important contributor to the narrow wavelength interval around 5200\,\AA\ where the $\Delta a$ system is quite sensitive, while the ${\Delta(V_1-G)}$ and especially $Z$ system are almost unaffected by Cr overabundance. But even with overabundant Cr the sensitivity of the $\Delta a$ system on magnetic field strength is the worst among other peculiar indices for low effective temperature.

The relation between values of peculiar parameters and the magnetic field strength is not linear due to
the saturation effects for stronger fields (${B\gtrsim 10}$\,kG), except $\Delta a$ which appears to have approximately linear dependance on magnetic field strength from 5 to 40 kG for ${\tef=8000}$\,K.

Finally, we investigated the question of stellar atmosphere fundamentals parameters determination based on calculated synthetic photometric colors of the Str\"omgren system. We found that the model atmosphere parameters derived using
the photometric calibrations for normal stars are not far from their true values and are within the usual error bars. We also showed that the maximum change of profiles of the hydrogen Balmer lines in comparison to the non-magnetic case does not exceed 2.5\% of the continuum level for the magnetic field strength 40\,kG.

In our further research we plan to build model atmospheres of CP stars introducing a grid of 1-D models on the stellar surface with individual values and orientations of magnetic field vector using technique described in this paper. This will allow us to study anisotropy effects induced by the magnetic field on line blanketing.

\begin{acknowledgements}
We are grateful to Prof. J. D. Landstreet for his very useful comments and corrections on the final version of the manuscript and Prof. V. Tsymbal for his permanent inspiration.
This work was supported by a Postdoctoral fellowship to SK at UWO funded by a Natural Science and Engineering Council of Canada Discovery Grant, and by INTAS grant 03-55-652 to DS. We also acknowledge support by the Austrian Science Fonds (FWF-P17890).
\end{acknowledgements}


\begin{thebibliography}{10}
\bibitem[Auer et al., 1977]{ahh}Auer, L. H., Heasley, J. N., \& House, L. L. 1977, \apj, 216, 531
\bibitem[Cramer \& Maeder, 1979]{cramer1}Cramer, N., \& Maeder, A. 1979, \aap, 78, 305
\bibitem[Cramer \& Maeder, 1980]{cramer2}Cramer, N., \& Maeder, A. 1980, \aap, 88, 135
\bibitem[del Toro Iniesta, 2003]{toro}del Toro Iniesta, J. S. 2003, Introduction to Spectropolarimetry (Cambridge University Press)
\bibitem[Hauck, 1974]{hauck}Hauck, B. 1974, \aap, 32, 447
\bibitem[Khan, 2004]{khan_synthm}Khan, S. A. 2004, \jqsrt, 88, 71
\bibitem[Khan et al., 2004]{khan_paper1}Khan, S. A., Kochukhov, O., \& Shulyak, D. 2004, in The A-Star Puzzle (UK: Cambridge University Press), Proc. IAU Symp., 224, 29
\bibitem[Kochukhov et al., 2005]{paper1}Kochukhov, O., Khan, S., \& Shulyak, D. 2005, \aap, 433, 671 (Paper I)
\bibitem[Kodaira, 1969]{kodaira}Kodaira, K. 1969, \apj, 157, 59
\bibitem[Kupka et al., 1999]{vald}Kupka, F., Piskunov, N., Ryabchikova, T. A., Stempels, H. C., \& Weiss, W. W. 1999, \aaps, 138, 119
\bibitem[Kupka et al., 2003]{kupka1}Kupka, F., Paunzen, E., \& Maitzen, H. M. 2003, \mnras, 341, 849
\bibitem[Kurucz, 1970]{kurucz_atlas9}Kurucz, R. L. 1970, SAO Special Report, 308
\bibitem[Kurucz, 1993a]{kurucz13}Kurucz, R. L. 1993, Kurucz CD-ROM 13, Cambridge, SAO
\bibitem[Kurucz, 1993b]{atlas12}Kurucz, R. L. 1993, in Peculiar versus Normal
Phenomena in A-type and Related Stars, ed. M. Dworetsky, F. Castelli, \& R. Faraggiana, Proc. IAU Coll., 138, ASP Conf. Ser., 44, 87
\bibitem[Leckrone, 1973]{leckrone1}Leckrone, D. 1973, \apj, 185, 577
\bibitem[Maitzen, 1976]{maitzen}Maitzen, H. 1976, \aap, 51, 223
\bibitem[Napiwotski et al., 1993]{napiwotski}Napiwotski, R., Sch\"onberner, D., \& Wenske, V. 1993, \aap, 268, 653
\bibitem[Piskunov, 1992]{piskunov_synth}Piskunov, N. 1992, in Stellar Magnetism, ed. Yu. V. Glagolevskij, \& I. I. Romanyuk (St. Petersburg: Nauka), 92
\bibitem[Piskunov \& Kochukhov, 2002]{piskunov}Piskunov, N., \& Kochukhov, O. 2002, \aap, 381, 736
\bibitem[Rees, 1987]{rees_nrt1}Rees, D. E. 1987, Numerical Radiative Transfer, ed. W. Kalkofen (Cambridge University Press), 213
\bibitem[Rees \& Murphy, 1987]{rees_nrt2}Rees, D. E., \& Murphy, G. A. 1987, Numerical Radiative Transfer, ed. W. Kalkofen (Cambridge University Press), 241
\bibitem[Rees et al., 1989]{rees}Rees, D. E., Murphy, G. A., \& Durrant, C. J. 1989, \apj, 339, 1093
\bibitem[Rogers, 1995]{templogg}Rogers, N. Y. 1995, Comm. Asteroseismol., 78
\bibitem[Shulyak et al., 2004]{llmodels}Shulyak, D., Tsymbal, V., Ryabchikova, T., St\"utz\, Ch., \& Weiss, W. 2004, \aap, 428, 993
\bibitem[Sobelman, 1977]{sobelman}Sobelman, I. I. 1977, Introduction to the theory of atomic spectra, Moscow
\bibitem[Socas-Navarro et al., 2000]{socas}Socas-Navarro, H., Trujillo Bueno, J., \& Ruiz Cobo, B. 2000, \apj, 530, 977
\bibitem[Stenflo, 1994]{stenflo}Stenflo, J. O. 1994, Solar Magnetic Fields (Dordrecht: Kluwer Academic Publishers)
\bibitem[Tsymbal, 1996]{tsymbal}Tsymbal, V. V. 1996, in Model Atmospheres and Spectral Synthesis, ed. S. J. Adelman, F. Kupka \& W. W. Weiss, ASP Conf. Ser., 108, 198
\bibitem[Tsymbal \& Shulyak, 2003]{tsymbal_ll}Tsymbal, V., \& Shulyak, D. 2003, in Modelling of Stellar Atmospheres, ed. N. Piskunov, W. W. Weiss, \& D. F. Gray, Astron. Soc. of the Pacific, CD-A14, Proc. IAU Symp., 210.
\end{thebibliography}
\end{document}